\documentclass[reprint,amsmath,amssymb,aps]{revtex4-1}

\usepackage{graphicx}
\usepackage{dcolumn}
\usepackage{bm}
\usepackage{hyperref}
\usepackage[utf8]{inputenc}
\usepackage{slashed}
\usepackage{mathtools}
\usepackage{makecell}
\usepackage[flushleft]{threeparttable}
\usepackage[english]{babel}
\usepackage{bm}
\usepackage[mathlines]{lineno}
\usepackage{color}
\definecolor{rossoCP3}{cmyk}{0,.88,.77,.40}
\definecolor{darkred}{rgb}{0.6,0,0}
\definecolor{drkgrn}{RGB}{0, 51, 0}
\hypersetup{
     colorlinks   = true,
     citecolor    = blue,
     urlcolor     = magenta,
      linkcolor    = darkred
}

\usepackage{listings}
\usepackage{color,xcolor}

\newcommand{\be}{\begin{equation}}

\newcommand{\ee}{\end{equation}}
\newcommand{\bea}{\begin{eqnarray}}
\newcommand{\eea}{\end{eqnarray}}

\newcolumntype{C}[1]{>{\centering\let\newline\\\arraybackslash\hspace{0pt}}m{#1}}

\lstdefinestyle{python}{
  belowcaptionskip=1\baselineskip,
  breaklines=true,
  frame=L,
  xleftmargin=\parindent,
  language=Python,
  showstringspaces=false,
  basicstyle=\small\ttfamily,
  morekeywords={models, lambda, forms,True,False,None},
  keywordstyle=\bfseries\color{deepgreen!40!black},
  commentstyle=\itshape\color{gray},
  identifierstyle=\color{black},
  stringstyle=\color{deepred},
  rulecolor=\color{gray},
}

\begin{document}

\title{Light new physics and neutrino electromagnetic interactions in XENONnT}

\author{Amir N.\ Khan}
\email{amir.khan@mpi-hd.mpg.de}
\affiliation{Max-Planck-Institut f\"{u}r Kernphysik, Postfach 103980, D-69029
Heidelberg, Germany}

\begin{abstract}
\noindent
We derive new limits on the neutrino electromagnetic interactions and weakly coupled light vector and scalar mediators using the recent XENONnT data of the solar neutrino-electron elastic scattering. XENONnT has already reported the world's best constraint on the flavor-independent effective neutrino magnetic moment with almost twice the exposure and improved systematics. We extend this analysis and derive constraints on all the possible electromagnetic interactions and flavor universal light gauge boson couplings and masses, which could contribute to the neutrino-electron elastic scattering process. We consider both flavor-independent and flavor-dependent interactions of the neutrino magnetic moments, millicharges, charge radii, and anapole moments for the electromagnetic interactions. The new limits on the magnetic moment, millicharge, vector, and scalar interactions are improved by about one order of magnitude. At the same time, there is relatively weaker improvement in the case of neutrino charge radii and anapole moments.
\end{abstract}

\date{\today}
\pacs{xxxxx}

\maketitle

\section{Introduction}
The XENONnT experiment has collected new data with a large size detector, the total exposure of $~1.16$ ton-years, reduced systematic uncertainties, and improved background \cite{Aprile:2022vux}. More importantly, the electronic recoil has been observed with no excess in the range $(1-7)$ keV like its predecessor detector XENON1T \cite{Aprile:2020tmw}. Furthermore, more than 50\% background reduction has been achieved in the new upgrade \cite{Aprile:2022vux}. On the one hand, the more extensive exposure and lower background rate of the experiment increase its sensitivity to dark matter detection. On the other hand, it can also increase its sensitivity to the new physics related to the solar neutrino interactions.

With this motivation in mind, we use the XENONnT new data and derive limits on all possible new interactions related to the massive neutrinos. These include electromagnetic interactions such as neutrino magnetic moment \citep{Fujikawa:1980yx, Shrock:1982sc, Vogel:1989iv,Abak:1989kp, Grimus:1997aa}, neutrino millicharge \cite{Babu:1989tq,Babu:1989ex,Foot:1990uf,Foot:1992ui,Davidson:1991si, Babu:1993yh,Bressi:2011yfa, Gninenko:2006fi,Chen:2014dsa,TEXONO:2018nir,Khan:2019cvi,Khan:2020vaf,Barbiellini:1987zz, Raffelt:1999gv, Davidson:2000hf, Melchiorri:2007sq, Studenikin:2012vi}, neutrino charge radius \cite{Bernabeu:2000hf, Bernabeu:2002nw, Bernabeu:2002pd, Fujikawa:2003ww}, anapole moment \cite{zel1958electromagnetic, zel1960effect, Barroso:1984re, Abak:1987nh, Musolf:1990sa,Dubovik:1996gx, Rosado:1999yn, Novales-Sanchez:2013rav} and new light vector and scalar mediators that feebly couple to neutrinos and electrons \cite{Fayet:1977yc,Boehm:2004uq,Langacker:2008yv,Paschos:2021hyb,Fayet:2020bmb}. We consider both flavor-independent effective interactions for the electromagnetic interactions by taking the same couplings for the three neutrino flavors contained in the final solar fluxes and flavor-dependent interactions. In the case of the weakly coupled new light mediators, we only consider the flavor universal and diagonal interactions.

Solar neutrino detectors offer a natural laboratory for testing the flavored new physics models related to the neutrino interactions because of the long baseline. The total flux thus contains all three flavors of neutrinos \cite{Khan:2017oxw,Amaral:2020tga,Coloma:2020gfv,Suliga:2020lir, Khan:2020csx,Khan:2022jnd}. The total rate is the sum of $\nu_e (\approx 56\%)$, $\nu_{\mu} (\approx 22\%)$ and $\nu_{\tau} (\approx 22\%)$ in case of the maximal mixing. In this case, the new interactions of muon and tau neutrinos are indistinguishable. However, in our analysis here, we will assume the non-maximal ``23'' mixing scheme, which makes it possible to distinguish, at least theoretically, between the muon and tau neutrino interactions. This effect has been included in eq. (\ref{eq:evnrate}). The flavor-dependent constraints become important when one compares the bounds with the flavor-specific experiments \cite{Deniz:2009mu, Borexino:2017fbd, Khan:2019cvi, Lindner:2018kjo}.

After the observation of the recoil electron excess by XENON1T \cite{Aprile:2020tmw}, several dark matter candidates and different neutrino nonstandard interactions were used to explain it, while several other studies derived new limits on model-dependent parameters \cite{Khan:2020vaf,Boehm:2020ltd, AristizabalSierra:2020edu, Okada:2020evk, Lindner:2020kko, Alonso-Alvarez:2020cdv, Chala:2020pbn, Ge:2020jfn, Amaral:2020tga, Benakli:2020vng, Chigusa:2020bgq, Li:2020naa, Baek:2020owl, Gao:2020wfr, Ko:2020gdg, An:2020tcg, McKeen:2020vpf, Bloch:2020uzh, Budnik:2020nwz,Farzan:2020dds,Khruschov:2020cnf,Kim:2020aua, Bally:2020yid, Arcadi:2020zni, Okada:2020evk,Li:2020naa, Shoemaker:2020kji,Khruschov:2020cnf,Shakeri:2020wvk,Takahashi:2020bpq, Takahashi:2020uio}. In principle, one can update all those results using the new XENONnT data. However, here we will focus only on the neutrino electromagnetic interactions and the weakly coupled new light mediator that could modify the neutrino-electron elastic scattering at low energy. We will treat the total predicted background reported by XENONnT by excluding the solar neutrino contribution. Then we will calculate the predicted neutrino spectrum in the presence of all the new physics interactions and add them to the background spectrum. We will use this setup to derive the one parameter at-a-time limits and the two parameters allowed parameter spaces.

After setting up the formal structure in section \ref{secII}, we discuss the analysis in section \ref{secIII}. In section \ref{secIV}, we present our results and conclude in section \ref{secV}.

\section{Formalism} \label{secII}


\begin{figure*}[tp]
\begin{center}
\includegraphics[width=5in, height=3in]{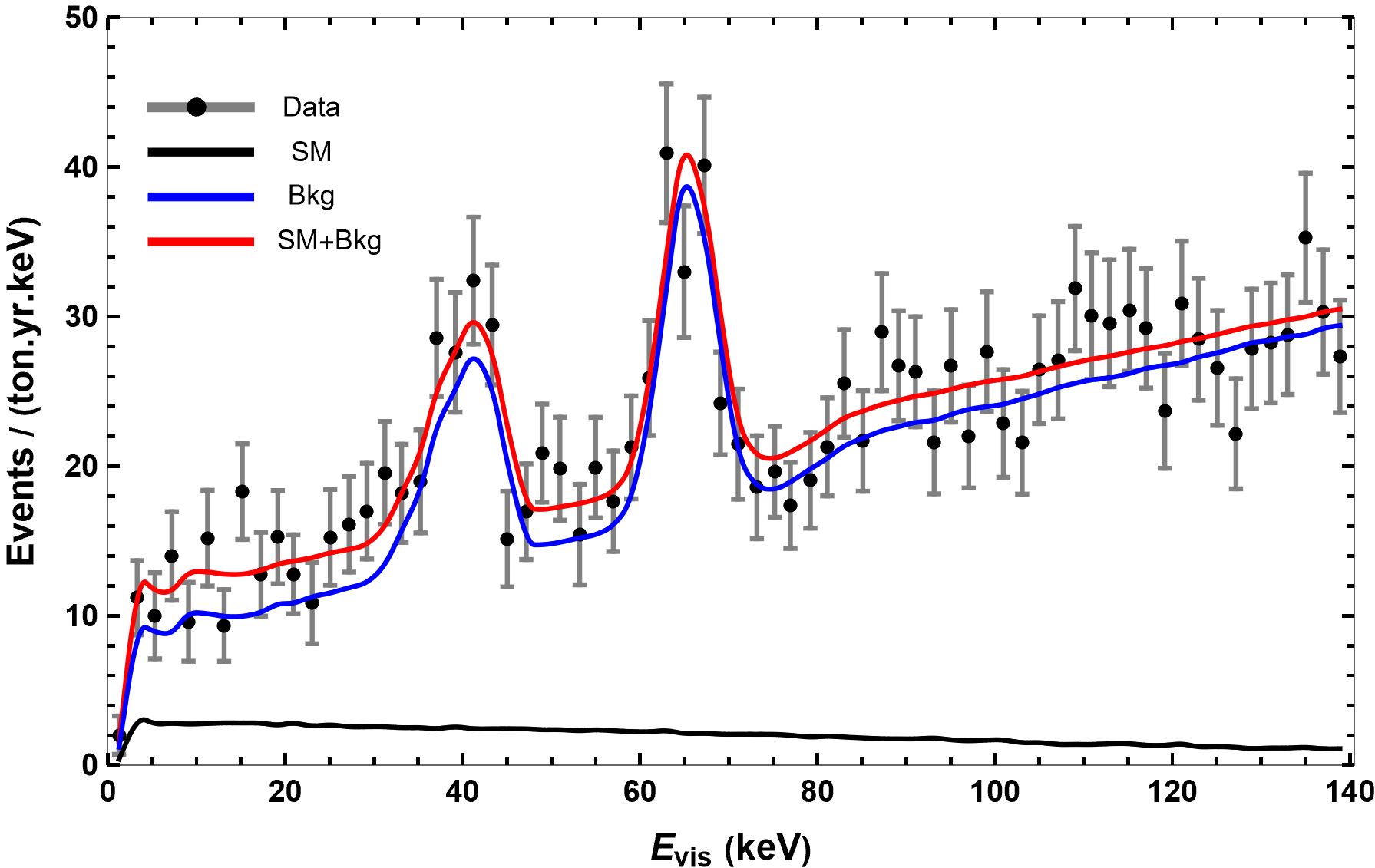}
\label{fig: spect}
\end{center}
\caption{{}\textbf{\ } Data, total background (red), background without solar neutrino contribution (blue) and our prediction for the solar neutrino energy spectrum (black). We obtain the background without the solar neutrino contribution (blue) by subtracting our calculated solar neutrino rates in each energy bin from the total background in the corresponding bin. We take the data and total background from ref. \cite{Aprile:2022vux}.}
\label{spectrum140}
\end{figure*}

Here we discuss the formalism needed for calculating expected rates due to the electromagnetic interactions and for light gauge bosons that mediate in the elastic neutrino-electron scattering process. The electroweak cross-section of the $\nu_{\alpha}-e$ scattering process in the standard model is given by

\begin{equation}
\left( \frac{d\sigma _{\nu _{\alpha }e}}{dE_{r}}\right) _{SM}=\frac{%
2G_{F}^{2}m_{e}}{\pi }\left[g_{L}^{2}+g_{R}^{2}\left( 1-\frac{E_{r}}{E_{\nu }}%
\right) ^{2}-g_{L}g_{R}\frac{m_{e}E_{r}}{E_{\nu }^{2}}\right],
\label{eq:xsecSM}
\end{equation}
where $G_{F}$ is Fermi constant, $m_e$ is mass of electron, $g_{L}=(g_{V}+
g_{A})/2$+1 for $\nu_{e}$, $(g_{V}+
g_{A})/2$ for $\nu_{\mu,\tau}$, $g_{R}=(g_{V}-g_{A})/2$ for $\nu_{e, \mu,\tau}$ , $g_{V}=-1/2+2\sin ^{2}\theta_{W}$, $g_{A}=-1/2$, $E_{\nu}$ is the incoming neutrino energy and $E_{r}$ is the electron recoil energy in the detector. We take $\sin ^{2}\theta_{W} =0.23867\pm 0.00016$ in the $\overline{\text{MS}}$ scheme \cite{ParticleDataGroup:2020ssz} and correct $g_{V}$ and $g_{A}$ for the small radiative corrections.

\subsection{Electromagnetic Interactions}
Among the four types of electromagnetic interactions, only the neutrino magnetic moment cross-section, as in the following, is added to the standard model cross-section (\ref{eq:xsecSM}) without any interference effects because of the chirality flipping of the neutrinos in the scattering process \citep{Fujikawa:1980yx, Vogel:1989iv, Dvornikov:2003js, Dvornikov:2004sj}, 

\begin{equation}
\left(\frac{d\sigma _{\nu _{\alpha }e}}{dE_{r}}\right) _{_{MM}}=\frac{\pi
\alpha_{em}^{2} \mu _{\nu_{\alpha} }^{2}}{m_{e}^{2}}\left[\frac{1}{E_{r}}-\frac{1}{%
E_{\nu }}\right],
\label{eq:mm}
\end{equation}
where $\alpha _{em}$ is the electromagnetic fine structure constant and $\mu _{\nu_{\alpha}}$ is the magnetic moment coupling in units of Bohr magneton ($\mu_B$).
We consider the neutrino flavor conserving cases for all types of electromagnetic interactions. For the millicharge neutrinos, charge radius, and anapole moment, we replace $g_{V}$ with $\widetilde{g}_{V}$, where
\begin{equation}
\widetilde{g}_{V}=g_{V}\ +\frac{\sqrt{2}\pi \alpha }{G_{F}}\left(\frac{%
\left \langle r_{\nu _{\alpha }}^{2}\right \rangle }{3}-\frac{a_{\nu _{\alpha }}}{18}-\frac{q_{\nu _{\alpha
}}}{m_{e}E_{r}}\right).
\label{eq:gvmod}
\end{equation}%
Here, $q_{\nu _{\alpha}}$ is the neutrino fractional electric charge in units of
unit charge of electron``e'' and $\left \langle r_{\nu _{\alpha }}^{2}\right \rangle$ and $a_{\nu _{\alpha }}$ are respectively the neutrino charge radius and anapole moment in units of $\rm cm^{2}$. Neutrino millicharge appears more sensitive among the three quantities because of its inverse squared dependence on electronic recoils. We will discuss this aspect in detail in the following sections.

\subsection{Light Mediators}
We will derive constraints on the coupling and masses of new light vector and scalar gauge bosons. These can contribute to the neutrino-electron elastic scattering process through the general model-independent vector (V), axial-vector (A), scalar (S), and pseudoscalar (P) interactions. At low recoils, since the new physics effects are inversely proportional to the recoil electron kinetic energy and the better agreement between the expected and observed background, XENONnT data can put stronger or competitive limits on the light mediator couplings and their masses. These interactions are predicted by a wide variety of models \cite{Fayet:1977yc,Boehm:2004uq,Langacker:2008yv,Paschos:2021hyb,Fayet:2020bmb}, however, we do not discuss here the possible origin of these interactions. An important aspect of such mediators is that they have low masses and very weak couplings. Therefore, their masses could be generated by spontaneous symmetry breaking well below the electroweak breaking scale. For other phenomenological implications of such interactions, see refs. \cite{Pospelov:2011ha,Pospelov:2012gm,Harnik:2012ni}.

In the following, we assume additional model independent light Spin-1 $(Z^{\prime}_{\mu})$ and spin-0 $(S)$ mediators which couple to electrons and the three flavors of neutrinos with equal coupling strengths via a vector, axial-vector, scalar, and pseudoscalar interactions. The following Lagrangians describe such interactions:
\begin{align}
\mathcal{L}_{V^{\prime}}&=\   
 -g_{V}^{^{\prime}}\left[\overline{\nu }_{L}\gamma
^{\mu }\nu _{L}+\overline{e}\gamma
^{\mu }e\right]Z^{^{\prime}}_{\mu} \ \ \ (\rm Vector),\\[10pt]
\mathcal{L}_{A^{\prime}}&=\  
 -g_{A}^{\prime}\left[\overline{\nu }_{L}\gamma
^{\mu }\gamma^{5}\nu _{L}+\overline{e}\gamma
^{\mu }\gamma^{5}e\right]Z^{^{\prime}}_{\mu } \ (\rm Axial-vector), \\[10pt]
\mathcal{L}_{S}&=\ -g_{S}\left[\overline{\nu }_{R}\nu_{L}+\overline{e}e\right]S+h.c  \ \ \ \ \ (\rm Scalar),\\[10pt]
\mathcal{L}_{P}&=\ -g_{P}\left[\overline{\nu }_{R}\nu
_{L}+\overline{e}\gamma^{5}e\right]S+h.c \  (\rm Pseudo-scalar).
\label{lagrangians}
\end{align}
Here $g_{V}^{^{\prime}}, g_{A}^{^{\prime}}, g_{A}$ and $g_{P}$ are the coupling constants of the corresponding interactions. Since the vector and axial-vector interactions interfere with the standard model interactions, in the low momentum transfer limit, the standard model couplings with electrons, $g_{V/A}$ in eq. $(\ref{eq:xsecSM})$ can be replaced by the effective parameters $\widetilde{g}_{V/A}$ \citep{Lindner:2018kjo,Khan:2020csx} as
\begin{align}
\widetilde{g}_{V} = g_{V}+\left( \frac{g_{V^{^{\prime }}}^{2}}{\sqrt{2}%
G_{F}(2m_{e}E_{r}+m_{V^{^{\prime }}}^2)}\right),
\label{gvpr}
\end{align}

and

\begin{align}
\widetilde{g}_{A} = g_{A}+\left( \frac{g_{A^{^{\prime }}}^{2}}{\sqrt{2}%
G_{F}(2m_{e}E_{r}+m_{A^{^{\prime }}}^2)}\right),
\label{gapr}
\end{align}
where $g_{V^{^{\prime }}}$ and $g_{A^{^{\prime }}}$ are the coupling constants and $m_{V^{^{\prime}}}$ and $m_{A^{^{\prime }}}$ are the respective masses of the new vector and axial-vector mediators.

The contribution of scalar mediators is added without interference. In this case, the scalar and pseudo-scalar interaction cross-sections  \citep{Cerdeno:2016sfi,Khan:2020csx} are%
\begin{align}
\left( \frac{d\sigma _{\nu _{\alpha }e}}{dE_{r}}\right) _{_{S}}=\left( 
\frac{g_{S}^{4}}{4\pi (2m_{e}E_{r}+m_{S}^2)^{2}}\right) \frac{%
m_{e}^{2}E_{r}}{E_{\nu }^{2}} \label{gs}, \\ \
\left( \frac{d\sigma _{\nu _{\alpha }e}}{dE_{r}}\right) _{_{P}}=\left( 
\frac{g_{P}^{4}}{8\pi (2m_{e}E_{r}+m_{P}^2)^{2}}\right) \frac{%
m_{e}E_{r}^2}{E_{\nu }^{2}},
\label{gp}
\end{align}
where $g_{S}$ and $g_{P}$ are the scalar and pseudoscalar coupling constants, and $m_{S }$ and $m_{P }$ are, respectively, their masses.

Notice that we do not consider the flavor-dependent light new interactions, although this is possible with the XENONnT of the solar neutrinos. It is an important direction to pursue because one can test different flavored additional vector type $U(1)^{\prime}$ models with data. We leave this for future work.
\subsection{Expected Energy Spectrum}
Next, we define the differential event rates as a function of the visible
recoil energy $(E_{vis})$ of electrons to estimate the electromagnetic and the new weak interactions that contribute to the standard model interactions. It can be
written as%
\begin{widetext}
\begin{equation}
\frac{dN}{dE_{vis}} = N_{e}\int_{E_{r}^{th}}^{E_{r}^{mx}}dE_{r} \int_{E_{\nu }^{mn}}^{E_{\nu }^{mx}}dE_{\nu
}\left( \frac{d\sigma _{\nu _{e}e}}{%
dE_{r}}\overline{P}_{ee}^{m} +\cos^2{\theta_{23}} \frac{d\sigma _{\nu _{\mu }e}}{dE_{r}}\overline{P}_{e\mu}^{m} +\sin^2{\theta_{23}} \frac{d\sigma _{\nu _{\tau}e}}{dE_{r}}\overline{P}_{e\tau}^{m}\right) \frac{d\phi }{dE_{\nu }}\epsilon (E_{vis})G(E_{vis},E_{r}),
\label{eq:evnrate}
\end{equation}
\end{widetext}
where $G(E_{r}, E_{vis})$ is a normalized Gaussian smearing function to
account for the detector finite energy resolution with resolution power $%
\sigma (E_{vis})/E_{vis}=(0.3171/\sqrt{E_{vis}[\text{keV}]})+0.0015$ and $\epsilon
(E_{vis})$ is the detector efficiency both taken from \cite{Aprile:2022vux},$%
\ d\phi /dE_{\nu }$ is the solar flux spectra were taken from \cite{Bahcall:2004mz} and $%
N_{e}$ is 1.16 ton-year exposure \cite{Aprile:2022vux}. Here, $d\sigma _{v_{\alpha }e}/dE_{r}$ are cross-sections, $\overline{P}_{ee}^{m}$ $\ $ and $\overline{P}_{e\mu /\tau }^{m}$ are the neutrino oscillation length averaged
survival ($\nu_e$) and conversion ($\nu_\mu, \nu_\tau$) probabilities of solar neutrinos in the presence of small matter effects
as given by,
\begin{equation}
\overline{P}_{ee}^{m}=s_{13}^{4}+\frac{1}{2}c_{13}^{4}{}(1+\cos 2\theta _{12}^{m}\cos
2\theta _{12})
\end{equation}%
and $\overline{P}_{e\mu/\tau}^{m}=1-\overline{P}_{ee}^{m}$ where s$_{ij}$, c$_{ij}$ are mixing
angles in vacuum and $\theta _{12}^{m}$ is the matter effects induced mixing
angle which was taken from \cite{Lopes:2013nfa, ParticleDataGroup:2020ssz}. We take values of oscillation parameters and their uncertainties from  \cite{ParticleDataGroup:2020ssz}, and we consider only the normal ordering scheme for the analysis. 
The integration limits are $E_{\nu
}^{mn}=(E_{r}+\sqrt{2m_{e}E_{r}+E_{r}^2})/2$ and $E_{\nu }^{mx}$ is the upper limit of each component of the pp-chain and CNO solar neutrinos considered here. We note that pp neutrinos are the dominant contributors to the energy range of interest here. At the same time, the other sources have a negligibly small effect on the energy spectrum due to the lower fluxes.
$E_{r}^{th}=1\ $keV is the detector threshold and $E_{r}^{mx}=140$ keV is the maximum electronic recoil energy of the experimental region of interest. Here, we note that the main contribution at the low energy recoils comes from the PP chain of solar neutrinos because of the large flux and energy up to about 1.3 MeV or below. Also, above the recoil energy of about 30 keV, there is a drastic increase in the total background and suppression of the solar neutrino signal. These aspects make the low energy end of the recoil spectrum more significant for the new interactions of solar neutrinos. 

\section{Analysis details} \label{secIII}

Having set out all the necessary formulas above, we calculate the
differential rate energy spectrum (eq. \ref{eq:evnrate}) as a function of $E_{vis}$ of the solar neutrinos using Eqs. (\ref{eq:xsecSM}, \ref{eq:gvmod} and \ref{eq:mm}) for the standard
and the neutrino electromagnetic interactions and Eqs. (\ref{eq:xsecSM}, \ref{gvpr} \ref{gapr}, \ref{gs} and \ref{gp}) for the vector and scalar interactions. 

We take the total background contribution below 140 keV from ref. \cite{Aprile:2022vux} and subtract our calculated energy spectrum from the total background. It
is shown in blue in fig. \ref{spectrum140}. Our calculated solar neutrino energy spectrum is shown in black in fig. \ref{spectrum140} which agrees well with the solar neutrino spectrum \cite{Aprile:2022vux}. The main background sources below 140 keV are $^{214}\rm Pb$, $^{85}\rm Kr$, $^{136}\rm Xe$, $^{124}\rm Xe$ and materials. 

It is important to note that we fit flavor-independent (effective) parameters for all four electromagnetic interactions following the limit reported by XENONnT on the effective magnetic moment. In addition, we derive limits on flavor-dependent parameters since XENONnT, like solar neutrino oscillation experiments, receives fluxes of all three neutrino flavors. It is clear from eq. (\ref{eq:evnrate}) where we assume the case of a non-maximal scheme of ``23'' mixing scheme. For the maximal mixing scheme, the bounds of muon and tau flavors would be the same. 

\begin{figure*}[tp]
\begin{center}
\includegraphics[width=7.2in, height=2.4in]{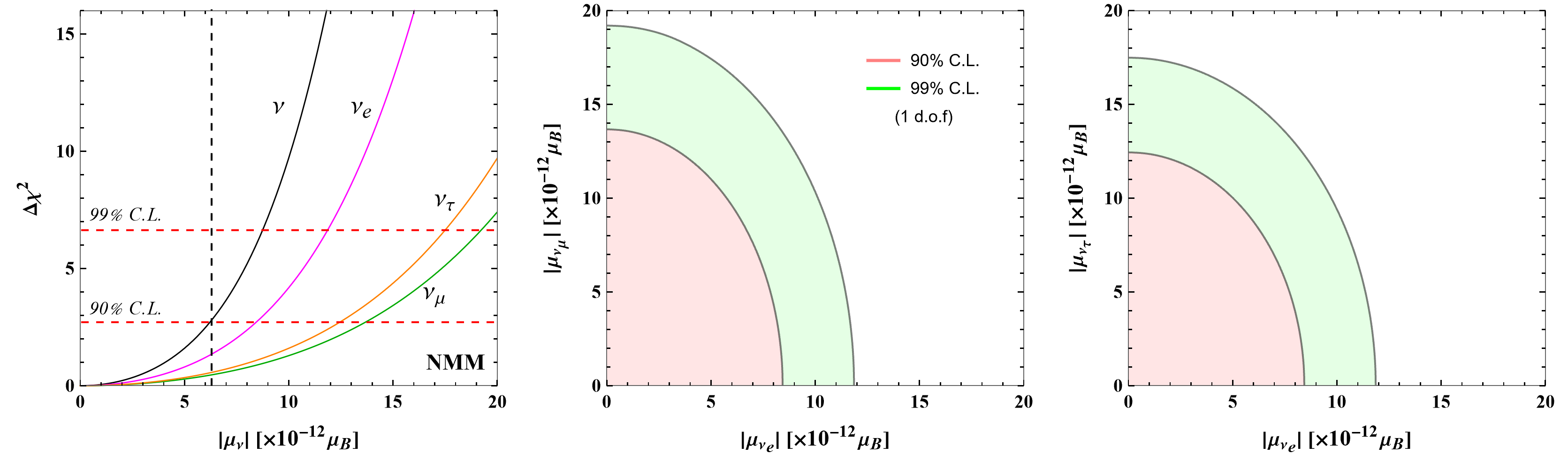}
\end{center}
\caption{{}\textbf{\ } One-dimensional $\Delta\chi^2$ distribution with 90\% and 99\% C.L. boundaries of neutrino magnetic moments (NMM) and two-dimensional allowed regions at 90\% and 99\% C.L. with one degree of freedom $\Delta\chi^2$. In the one-dimensional case, the distribution in black corresponds to the effective flavor-independent magnetic moment. All the other new physics parameters were set to zero when we fit the one or two parameters shown here.}   
\label{fig:NMM}
\end{figure*}

To estimate the new physics parameters using XENONnT data, we follow the least-squared statistical method and define the following $\chi^{2}$ function,

\begin{align}
\chi ^{2} & =\sum_{n=1}^{70} \left(\frac{%
(\frac{dN}{dE_{vis}}(1+\alpha)+B)^i_{th}-(\frac{dN}{dE_{vis}})_{obs}^{i}}{\sigma
^{i}}\right)^2 \notag \\
& \ \ \ \ \ \ \ \ \ + \left(\frac{\alpha}{\sigma_{\alpha}}\right)^2
\label{eq:chisq}
\end{align}%
where the expression in the bracket
(.....)$_{th}^{i}$ corresponds to the expected number of events in the $i-$th bin
which is the sum of the solar neutrinos rate and the background for each bin, while the bracket (...)$_{obs}^{i}$ corresponds to the observed number of events, and $\sigma ^{i}$ is the experimental uncertainty in the respective bin. We show the data point with uncertainties, the solar neutrino expected energy rates, and the background rates in fig. \ref{spectrum140}. We add the pull term to account for the theoretical uncertainty, mainly from the solar fluxes. We take $\alpha$ as the pull parameter with "$\sigma_{\alpha} = 10\%$" uncertainty in the fluxes \cite{Aprile:2022vux}. In addition, we also include penalty terms corresponding to the "$\theta_{12}$" "$\theta_{13}$" and "$\theta_{23}$" oscillation parameters to account for their uncertainties. 
We are using eq. (\ref{eq:chisq}), we reproduce the upper limit $\sim 6.26 \times 10^{-12}\mu _{B}$ at 90\% C.L. on the neutrino flavor independent (effective) magnetic moment, as shown in the left-hand side plot of fig. \ref{fig:NMM}, which is in the best agreement with the limit reported in ref. \cite{Aprile:2022vux}, that is, $\sim 6.3 \times 10^{-12}\mu _{B}$. This result is shown on the left-most in black in fig. \ref{fig:NMM}.

\begin{figure*}[tp]
\begin{center}
\includegraphics[width=7.2in, height=2.4in]{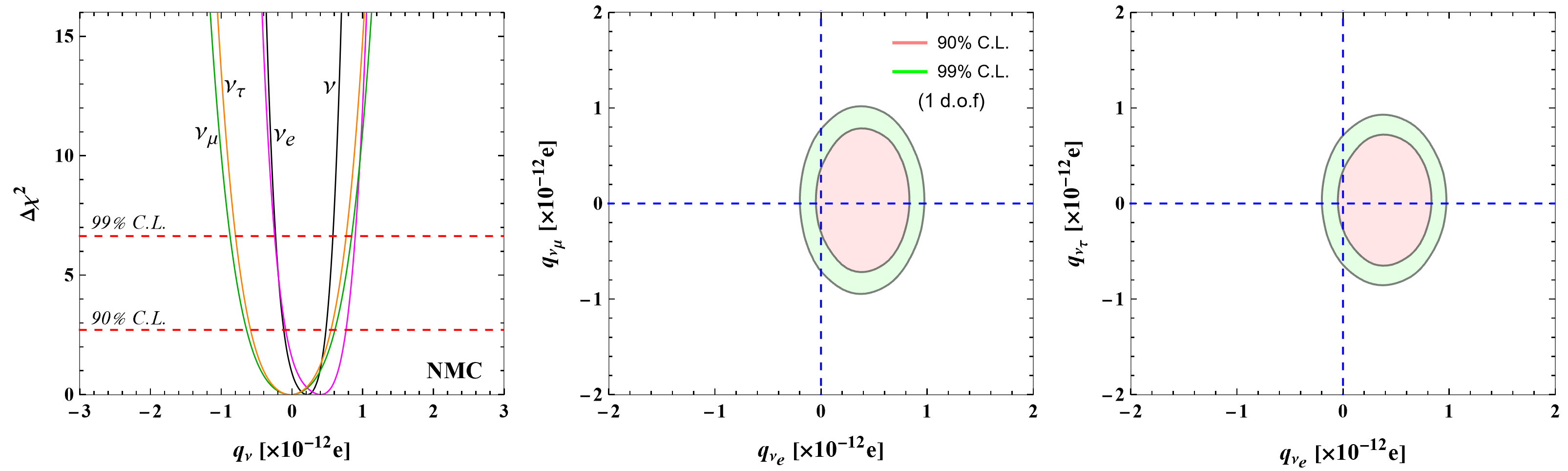}
\end{center}
\caption{{}\textbf{\ } One-dimensional $\Delta\chi^2$ distribution with 90\% and 99\% C.L. boundaries of neutrino millicharge (NMC) and two-dimensional allowed regions at 90\% and 99\% C.L. with one degree of freedom $\Delta\chi^2$. In the one-dimensional case, the distribution in black corresponds to the effective flavor-independent neutrino millicharge. All the other new physics parameters were set to zero when we fit the one or two parameters here.}   
\label{fig:NMC}
\end{figure*}

\begin{figure*}[tp]
\begin{center}
\includegraphics[width=7.2in, height=2.4in]{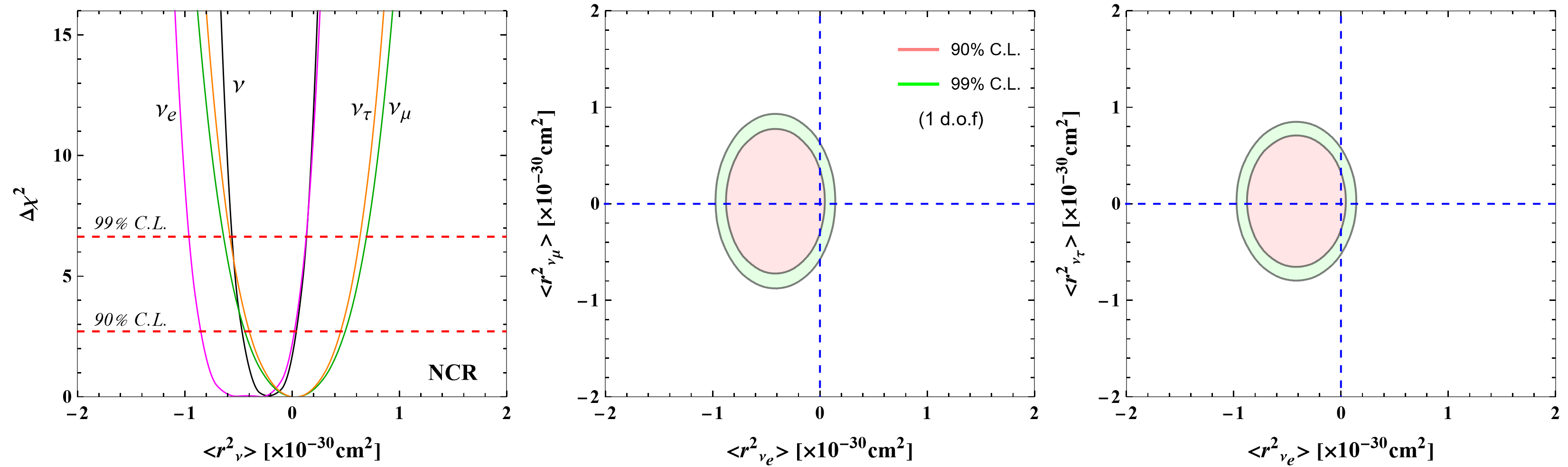}
\end{center}
\caption{{}\textbf{\ } One-dimensional $\Delta\chi^2$ distribution with 90\% and 99\% C.L. boundaries of neutrino charge radius (NCR) and two-dimensional allowed regions at 90\% and 99\% C.L. with one degree of freedom $\Delta\chi^2$. In the one-dimensional case, the distribution in black corresponds to the effective flavor-independent charge radius. All the other new physics parameters were set to zero when we fit the one or two parameters here.}   
\label{fig:NCR}
\end{figure*}

\begin{figure*}[tp]
\begin{center}
\includegraphics[width=7.2in, height=2.4in]{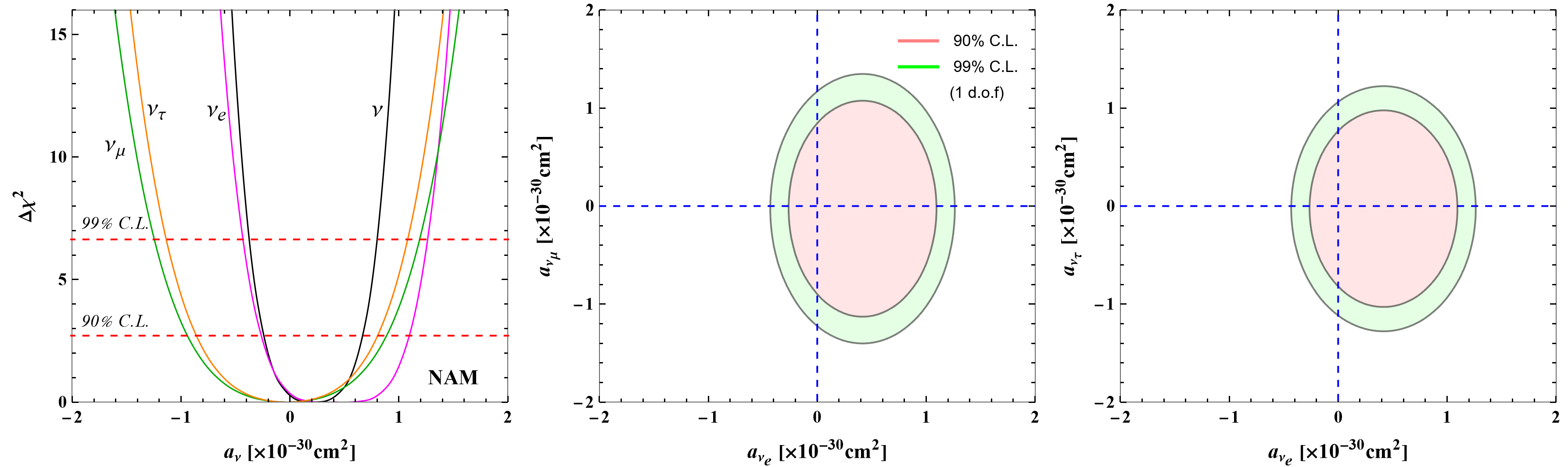}
\end{center}
\caption{{}\textbf{\ } One-dimensional $\Delta\chi^2$ distribution with 90\% and 99\% C.L. boundaries of neutrino anapole moment (NAM) and two-dimensional allowed regions at 90\% and 99\% C.L. with one degree of freedom $\Delta\chi^2$. In the one-dimensional case, the distribution in black corresponds to the effective flavor-independent anapole moment. All the other new physics parameters were set to zero when we fit the one or two parameters here.}    
\label{fig:NAM}
\end{figure*}

\begin{table*}[t]
\begin{center}
\begin{tabular}{c|c|c|c|c}
\hline \hline
Flavor & $|\mu _{\nu }|[\times 10^{-11}\mu _{B}]$ &  $q_{v}\ [\times 10^{-13}e]$& $\left \langle r_{\nu
}^{2}\right \rangle \ [\times 10^{-32}$cm$^{2}]$
& $a_{\nu }\ [\times 10^{-32}$cm$^{2}]$ 
\\ \hline 
$\nu\ ($XENONnT$)$ & $ <0.63 $ & $\  \ [-1.3,\ 4.7$ 
$]\ $& $\ [-45,\ 3.0$ $]$  & $\ [$\ $-23,65]$ 
\\ \hline
$\nu _{e}\ ($XENONnT$)$ & $ <0.85 $ & $\  \ [-2.5,\ 9.0$ 
$]\ $& $\ [-85,\ 2.0$ $]$  & $\ [$\ $-26,110]$ \\ 
$\nu _{\mu }($XENONnT$)$ & $ < 1.37$ & $[-8.9,\ 8.6$ $%
]\ $&  $\ [-45,\ 52$ $]$ & $[-95, 89]$ \\ 
$\nu _{\tau }($XENONnT$)$ & $ <1.24 $ & $[-7.9,\ 7.8$ 
$]\ $ & $\ [-40,\ 45$ $]$   & $[-86, 79]$ \\ \hline \hline
$\nu _{e}\ ($Others$)$ & \multicolumn{1}{|l|}{$%
\begin{array}{l}
\leq 3.9\  \text{(Borexino)}\  \\ 
\multicolumn{1}{c}{\leq 110\  \text{(LAMPF)}\ } \\ 
\leq 11\  \text{(Super-K)} \\ 
\multicolumn{1}{c}{\leq 7.4\  \text{(TEXONO)}} \\ 
\multicolumn{1}{c}{\leq 2.9\  \text{(GEMMA)}}%
\end{array}%
$} & $%
\begin{array}{l}
\leq 15 \\ 
\text{(Reactor)}%
\end{array}%
\ $  & $\  \ 
\begin{array}{l}
\lbrack 0.82,\ 1.27]\  \text{(Solar)} \\ 
\lbrack -5.94,\ 8.28]\  \text{(LSND)} \\ 
\lbrack -4.2,\ 6.6]\  \  \text{(TEXONO)}%
\end{array}%
$& $-$ \\ \hline
$\nu _{\mu }($Others$)$ & \multicolumn{1}{|l|}{$%
\begin{array}{c}
\leq 5.8\  \text{(Borexino)\ } \\ 
\leq 68\  \text{(LSND)\ } \\ 
\leq 74\  \text{(LAMPF)}[7]%
\end{array}%
$} & $-$ & \multicolumn{1}{|l|}{$%
\begin{array}{l}
\lbrack -9,\ 31]\  \text{(Solar)} \\ 
\leq 1.2\  \text{(CHARM-II)} \\ 
\lbrack -4.2,\ 0.48]\  \text{(TEXONO)}%
\end{array}%
$}  & $-$ \\ \hline
$\nu _{\tau }($Others$)$ & \multicolumn{1}{|l|}{$%
\begin{array}{l}
\leq 5.8\  \text{(Borexino)} \\ 
\multicolumn{1}{c}{\leq 3.9\times 10^{4}\  \text{(DONUT)}}%
\end{array}%
$} &  $%
\begin{array}{c}
\leq 10^{-8}\  \\ 
\text{(Beam dump)}%
\end{array}%
$  &$
\begin{array}{l}
\lbrack -9,\ 31]\  \text{(Solar)}%
\end{array}%
$  & $-$ \\ \hline \hline
\end{tabular}%
\\[0pt]
\end{center}
\caption{90\% C.L. bounds on neutrino magnetic moment, charge radius,
 millicharge and anapole moment from XENON1T and other
 laboratory experiments. The first row corresponds to the flavor-independent, effective parameters. For comparison with astrophysical
 constraints see ref. \cite{ Giunti:2015gga} and for COHERENT, see refs. \cite{Khan:2019cvi}. Apart from Borexino \cite{Borexino:2017fbd} and
 solar \cite{Khan:2017djo}, bounds from all other experiments
 were taken from refs. \cite{ Giunti:2015gga,Borexino:2017fbd}.}
\label{tabel1}
\end{table*}

\begin{figure*}[tp]
\begin{center}
\includegraphics[width=7in, height=7in]{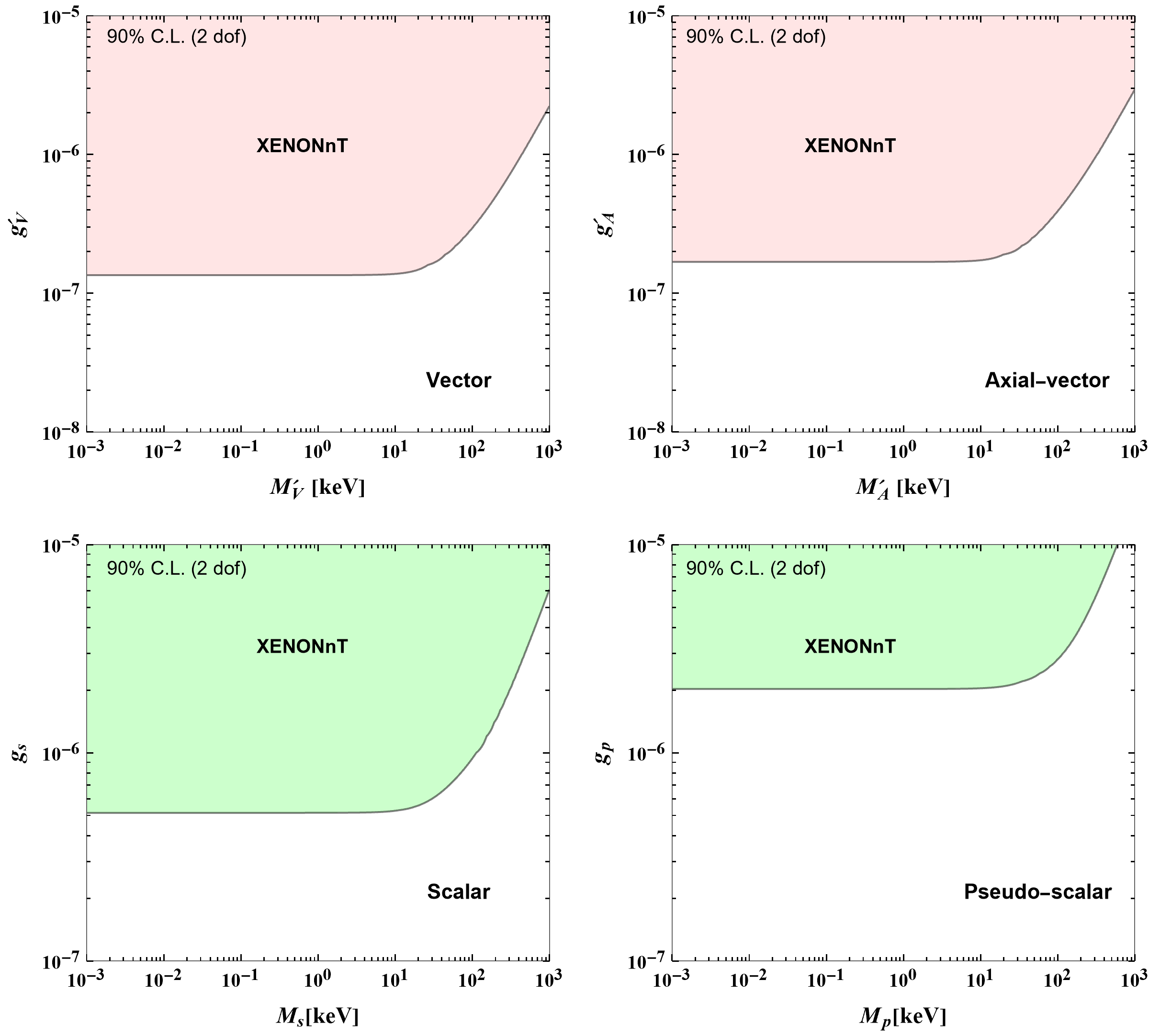}
\end{center}
\caption{The 90\% C.L. two d.o.f $\Delta \chi^2$ excluded regions in the parameter spaces of the light mediator masses and their couplings to neutrinos and electrons using the XENONnT new data.}
\label{2d_mediators}
\end{figure*}

\begin{table*}[t]
\begin{center}
\begin{tabular}{c|c|c|c|c|c}
\hline \hline
Coupling & XENONnT (\textbf{this work}) & PandaX-II & GEMMA & Borexino & TEXONO \\ \hline
$g_{V^{^{\prime }}}(\times 10^{-7})$ & $\lesssim 1.3 $  & $\lesssim 32$ & $\lesssim 5.0$  & $\lesssim 17$ & $\lesssim 58$ \\ \hline
$g_{A^{^{\prime }}}(\times 10^{-7})$ & $\lesssim 1.7 $  & $\lesssim 34$ & $-$ & $-$ & $-$ \\ \hline
$g_{S}\ (\times 10^{-7})$ & $\lesssim 5.0 $  & $\lesssim 49$ & $\lesssim 6.0$ & $\lesssim 6.0$ & $-$ \\ \hline
$g_{P}\ (\times 10^{-7})$ & $\lesssim 20 $  & $\lesssim 67$ & $-$ & $-$ & $-$ \\ \hline \hline
\end{tabular}%
\end{center}
\caption{90\% C.L. (2 dof) upper bounds on the coupling constants of four types of interactions considered in this work. These limits correspond to the mediator masses: $m_{V}^{\prime} {\lesssim 10}$ keV (vector), $m_{A}^{\prime}$ ${\lesssim 10}$ keV (axial-vector), ${m_{S}\lesssim 15}$ keV (scalar) and ${m_{P}\lesssim 30}$ keV (pseudoscalar) interactions. One can read these bounds also directly from fig. \ref{2d_mediators}. The limits for GEMMA  \cite{Beda:2009kx}, Borexino \cite{Agostini:2018uly}, TEXONO \cite{Deniz:2010mp}, PANDA-XII \cite{Khan:2020csx} are given for comparison. For comparison with the astrophysical limits, see ref. \cite{Harnik:2012ni}.}
\label{tableII}
\end{table*}

\section{Results and Discussion} \label{secIV}

Fig. \ref{fig:NMM} shows the results of our analysis for the flavor-dependent neutrino magnetic moments in one-dimensional and two-dimensional contour plots. In both cases, we show the 90\% and 99\% C.L. allowed regions. In the one-dimensional case, we retain one parameter and put the other two equal zero. In contrast, we fit two parameters in the two-dimensional case and put the third parameter equal zero. All the 90\% C.L. limits are summarized in the table's first three rows and 2nd column (\ref{tabel1}). We also show bounds from the other experiments in the table.

Next, we repeat the above exercise for the neutrino millicharges, charge radii, and neutrino anapole moments. We obtain the one-dimensional and two-dimensional allowed parameter spaces at 90\% and 99\% C.L., which are shown, respectively, in fig \ref{fig:NMC}, \ref{fig:NCR} and \ref{fig:NAM}. The bounds at 90\% C.L. are summarized in the table's first three rows and 3rd, 4th and 5th columns (\ref{tabel1}).

As shown in fig. \ref{fig:NMM}, \ref{fig:NMC}, \ref{fig:NCR} and \ref{fig:NAM}, we fit both flavor-independent and flavor-dependent parameters for each type of interaction. In the former case, we take the electromagnetic parameter as a common parameter for the three neutrino fluxes. In contrast, we take a separate parameter for each flux corresponding to its flavor in the latter case. This choice is motivated by the fact that because of the very long baselines, the solar neutrinos contain all three flavors and thus present a natural laboratory for any flavor-dependent new physics related to the neutrino interactions. The flavor-independent effective parameter bounds are stronger than the flavor-dependent bounds for all types of interactions because their contributions to the total rate are the same. It is evident from all four figures and table \ref{tabel1}. In the case of flavor-dependent interactions, the $\nu_{\mu}$ and $\nu_{\tau}$ flavors are distinguished through the atmospheric mixing angle, $\theta_{23}$, as given in eq. (\ref{eq:evnrate}).

Interestingly, among the four types of electromagnetic interactions, the neutrino magnetic moment and millicharge for the flavor-independent and the $\nu_{e}$ flavor cases get stronger limits about one order of magnitude than the previous bounds. For the other flavors, there is also a considerable improvement. Compared to the neutrino magnetic moment and milli-charge, the constraints on the charge radii and the neutrino anapole moments have relatively smaller improvements than the previous ones. It is because there is no inverse recoil energy dependence in latter cases, as evident from eq. (\ref{eq:gvmod}).

Next, we fit the couplings and masses of the four types of new weak interactions. Fig. \ref{2d_mediators} shows all the results in the form of two-dimensional exclusion plots at 90\% C.L.., and the corresponding bounds at 90\% C.L. are summarized in table (\ref{tableII}). The table clearly shows that limits are significantly improved in all cases. In particular, the vector, axial-vector, and scalar coupling limits get stronger by at least one order of magnitude and supersede the previous limits from the terrestrial experiments. The vector couplings get the strongest bounds, $g_{V^{^{\prime }}} \lesssim 1.3$ for the mediator mass of $m_{V^{^{\prime }}} \lesssim 10$ keV, the axial-vector coupling $g_{A^{^{\prime }}} \lesssim 1.7$ for the mediator mass of $m_{A^{^{\prime}}} \lesssim 10$ keV, the scalar coupling $g_{S} \lesssim 5$ for the mediator mass of $m_{S} \lesssim 15$ keV and the pseudo-scalar coupling $g_{P} \lesssim 20$ for the mediator mass of $m_{P} \lesssim 30$ keV, which are the best constraints so far. Note that these constraints also supersede the bounds from the Panda-XII experiment \cite{Khan:2020csx, Zhou:2020bvf}, which is a similar dark matter detector experiment to XENONnT.

\section{Summary and Conclusion} \label{secV}
We have analyzed the new data of XENONnT to constrain the electromagnetic interactions and new, feebly coupled, weak interactions. Unlike XENON1T, the new data has not revealed any excess at the low-energy electronic recoils, and the experiment has improved its systematics and background. With such an improvement, one can expect stronger constraints from the solar neutrino interactions. Therefore, we have derived limits on the neutrino electromagnetic interactions such as neutrino magnetic moment, millicharge, charge radius, and anapole moment. Further, we also derive limits on the weakly coupled vector, axial-vector, scalar, and pseudoscalar interactions. We have first reproduced the 90\% C.L. bound on the neutrino flavor-independent effective neutrino magnetic moment, which agrees well with the one reported by the XENONnT collaboration \cite{Aprile:2022vux}. Then, to better compare with other laboratory bounds on the flavor-dependent electromagnetic interactions, we have introduced both the effective flavor-independent and flavor-dependent parameters and have constrained them with the new data. All the results are shown in one-dimensional $\Delta \chi^2$ distributions and two-dimensional allowed parameter spaces at 90\% and 99\% C.L.. The 90 \% C.L. limits for all cases are given in table \ref{tabel1}. For a quick comparison, the bounds from other laboratory experiments and astrophysical observations are also given in the table.

Because of the enhanced sensitivity of the magnetic moment and the neutrino millicharge at lower electronic recoils, we obtain stronger constraints for both cases. However, the charge radius and anapole moment lack this characteristic, and therefore bounds obtained are only slightly improved. The inverse recoil electron energy dependence makes the low energy search data a unique probe of the neutrino magnetic moment and neutrino millicharge. The $\nu-e$ scattering constraints on the neutrino magnetic moment and neutrino millicharge are naturally stronger than those from the coherent elastic neutrino-nucleus scattering \cite{Khan:2022jnd, Khan:2022not, Coloma:2022umy} for the kinematical reasons. The small target electron mass scales up the enhanced sensitivity at the lower electronic recoils in the former case while the large nuclear mass suppresses this sensitivity in the latter case (see ref. \cite{Khan:2022not} for a detailed discussion about this issue). In contrast, in the latter case, it is scaled up by the target nuclear mass, which partially reduces the enhancement of low energy sensitivity. On the other hand, the new limits in the case of neutrino charge radii and neutrino anapole moments are still weaker than previous laboratory constraints. 

Comparison with other bounds shows that there is one order of magnitude improvement from other laboratory experiments for the case of magnetic moment and neutrino millicharges, as shown in table \ref{tabel1}. However, for the neutrino millicharges, the bound from the neutrality of matter is still eight orders of magnitude stronger than the new bound. Also, the neutrino millicharge constraints derived in this work are about one order of magnitude weaker than those from the astrophysical observations \cite{Raffelt:1999gv, Davidson:2000hf}. However, the future upgrades of XENONnT, LZ, Darkside-20k, and DARWIN \cite{Aalbers:2020gsn, Aalseth:2017fik, Aalbers:2016jon} are likely to compete with these constraints or directly detect these interactions.

In the case of the weakly coupled new light gauge bosons mediating the elastic neutrino-electron scattering process, we consider the flavor diagonal and flavor universal interactions. We derive the excluded regions between the respective couplings and masses. The results at 90\% C.L. are shown in fig. \ref{2d_mediators} and all the limits are summarized in table \ref{tableII}. The limits on the vector and scalar couplings improve by one order of magnitude for mediator mass below 15 keV for the vector interactions and below 30 keV for the scalar interactions.

To conclude, neutrino electromagnetic interactions and new weakly coupled vector and scalar interactions are predicted by several models beyond the standard model with massive neutrinos. The observation of tiny yet finite neutrino masses in the oscillation experiments makes such interactions likely in both dedicated neutrino experiments and the direct detection dark matter experiment like XENONnT through the low-energy scattering processes. Their effects on the astrophysical phenomenon are equally important. The stronger constraints obtained in this work, particularly for the neutrino magnetic moment, neutrino millicharges, and vector and scalar interactions, are an indication of the 
minimum sensitivity needed to push the bounds further to reveal direct observation of such interactions and rule out or support new physics models.

\begin{acknowledgments}
The author thanks Andrii Terliuk, Knut Dundas Moraa, and Jingqiang Ye for providing the necessary information about the XENONnT data. Alexander von Humboldt Foundation financially supports this work under the postdoctoral fellowship program.
\end{acknowledgments}

\bibliography{V1XENONnT}

\end{document}